\documentclass[manuscript]{aastex}

\pdfoutput=1

\shorttitle{Repetitive reconnections and flux-rope evolution}
\shortauthors{Kumar et al.}

\begin{document}

\title{On the role of repetitive magnetic reconnections in evolution of
magnetic flux-ropes in solar corona}

\author{Sanjay Kumar, R. Bhattacharyya, and Bhuwan Joshi} 
\affil{Udaipur Solar Observatory, Physical Research Laboratory, Dewali, Bari Road, Udaipur-313001, India}

\author{P. K. Smolarkiewicz}
\affil{European Centre for Medium-Range Weather Forecasts, Reading RG2 9AX, UK}


\begin{abstract}
Parker's magnetostatic theorem extended to astrophysical magnetofluids
with large magnetic Reynolds number supports ceaseless regeneration of
current sheets and hence, spontaneous magnetic reconnections recurring
in time. Consequently, a scenario is possible where the repeated
reconnections provide an autonomous mechanism governing emergence of
coherent structures in astrophysical magnetofluids. In this work, such
a scenario is explored by performing numerical computations commensurate
with the magnetostatic theorem. In particular, the computations explore
the evolution of a flux-rope governed by repeated reconnections
in a magnetic geometry resembling bipolar loops of solar corona. The
revealed morphology of the evolution process – including onset and
ascent of the rope, reconnection locations and the associated topology
of the magnetic field lines – agrees with observations, and thus
substantiates physical realisability of the advocated mechanism.
\end{abstract}


\section{Introduction}
The astrophysical plasmas in general, and the solar corona in particular,
are described by the non-diffusive limit of magnetohydrodynamics
(MHD). The reason for such description is the large length scales and
high temperatures, inherent to such plasmas which make the Lundquist
number ($S=v_aL/\eta$, in usual notations) extremely high. For
example, the Lundquist number for the solar corona, having a typical
length scale $L\approx 10^7$m and magnetic diffusivity $\eta\approx
1\rm{m}^2s^{-1}$ {\citep{aschwanden}}, is in the orders of $10^{13}$.
In such high-$S$ plasmas, the Alfv\'{e}n's  theorem of flux-freezing
{\citep{frozen}} is satisfied, thus predicting magnetic field lines
(MFLs) to remain tied with fluid parcels during an evolution. With
MFLs under coevolution with fluid parcels, a magnetic flux surface
(MFS)---made by the loci of MFLs---once identified with a fluid surface,
will maintain the identity throughout the evolution. As demonstrated
by recent numerical simulations {\citep{sanjay-pap, complexity-pap}},
such coevolution spontaneously develop current sheets (CSs); i.e., two
dimensional surfaces of intense volume current density across  which
the MFLs flip sign.  These simulations attribute the appearance of CSs
to favorable contortions of MFSs, generic to the high-$S$ magnetofluids
dynamics. Noteworthy in the simulations is the development of a quasi
steady state, approximately concurrent with the growing CSs. This is
expected as the favorable contortions bring anti-parallel field lines
in proximity and thereby increase local magnetic pressure which in turn,
opposes further contortions. In the ideal scenario of $\eta=0$, the
quasi steady state corresponds to a steady state where  magnetic field
is discontinuous across a CS.

The above findings are in conformity with Parker's magnetostatic theorem
{\citep{parker-72, parker-88, parker-book, parker-ppcf}}. The theorem
states that the development of CSs is ubiquitous in an equilibrium
magnetofluid with infinite electrical conductivity and complex
magnetic topology. The development is due to a general failure of
spatially continuous magnetic field in achieving local equilibrium
while preserving its topology. Generalization of the theorem to a
magnetofluid undergoing topology-preserving evolution then naturally
favors inevitable onset of CSs when the fluid relaxes toward a steady
state. Here, the inevitability is due to unbalanced forces, which under
flux-freezing have a constrained evolution that develop sharp gradients
in the magnetic field {\citep{parker-book}}.

In presence of an otherwise negligible magnetic diffusivity, the CSs
provide sites where the Lundquist number is locally reduced. The
flux-freezing is  destroyed and the locally diffusive magnetofluid
undergoes  magnetic reconnections (MRs) where magnetic energy is
converted into energy of mass flow and heat.  With reconnection, the
CSs are dissipated, and MFLs  being tied to fluid parcels are expunged
from the reconnection sites along with the mass flow.  These expunged
field lines push onto other MFLs and, under favorable conditions, may
lead to further reconnections. Importantly, a single reconnection
can initiate consecutive secondary MRs, intermittent in space and time,
 which may shape up the dynamics of high-$S$ magnetofluids. In a recent
computational study \citep{dinesh-2015} a scenario was explored, where
repeated reconnections were identified as a cause for  generating various
magnetic structures, some duplicating magnetic antics of the Sun. 

Toward realizing the above scenario in solar corona, we note that hard
X-ray coronal sources are standardly accepted as signatures of
magnetic reconnection in solar flares \citep{krucker2008}.  High
resolution measurements show multiple peaks of non-thermal nature
in flare's time profile, which suggest the  corresponding energy
releases and hence, the underlying reconnections, to be episodic
\citep{bhuwan2013}.  Recent observations at multiple channels (hard
X-ray and extreme ultraviolet) reveal intense localized  brightening
to occur below an ascending flux-rope \citep{kushwaha2015}.   The
spatio-temporal correlation between intermittent energy release and
ascent of an overlying flux-rope imply a causal connection between
magnetic reconnections and the ascent \citep{cho2009, cheng2011}.

In the above perspective, this paper numerically demonstrates the
evolution of a flux-rope in terms of its creation from initial bipolar
field lines and continuous ascent, mediated via  the common process
of repeated spontaneous reconnections. The corresponding manifestations 
of magnetic topology---in likes of MRs occurring at a height, reconnections
being localized below the rope and development of dips at the bottom part of
the rope---are in general harmony with observations. This contrasts 
with the contemporary simulations of flux-ropes that judiciously precondition 
rope evolution---either by specifying preexisting twisted magnetic 
structures that emerge from below the photosphere 
{\citep{fan2001,fan&gibson2003,fan2010,fan2011,chatterjee2013,fan2016}}, or
by forcing magnetic reconnections inside a sheared arcade with select 
(shearing and/or converging) initial flows
{\citep{van&martens,choe1996,amari1999,amari2003,aulanier2010,xia2014}}.
While the present work also explores MRs inside sheared arcades, the
computations being in agreement with the magnetostatic theorem generate 
spontaneous reconnections which provide an autonomous  mechanism
to govern dynamics of any high-$S$ magnetofluid.  The novelty of this work
is in its approach to establish creation and activation of a physically
realizable flux-rope as a consequence of the autonomous mechanism.
Introspectively, such a rope has the flavor of a self-organized state
\citep{kusano1994}.

To make numerical simulations agree with the theory of CS formation, the
requirement is the satisfaction of flux-freezing to high fidelity between
two successive MRs, while enabling local diffusion of MFLs co-located
and concurrent with the developing CSs. These two  requirements are
achieved through the following numerical scheme.  Viscous relaxation
\citep{low-bhattacharyya,sanjay-pap,dinesh-2015,complexity-pap} of
a thermally homogeneous,  incompressible magnetofluid with infinite
electrical conductivity is employed  to evolve the fluid under the
flux-freezing till CSs develop.   To focus on the idea, the relevant
Navier-Stokes MHD equations are
\begin{eqnarray}
\label{stokes}
& & \frac{\partial{\bf{v}}}{\partial t} 
+ \left({\bf{v}}\cdot\nabla \right) {\bf{ v}} =-\nabla p
+\left(\nabla\times{\bf{B}}\right) \times{\bf{B}}+\frac{\tau_a}{\tau_\nu}\nabla^2{\bf{v}} ~~~,\\  
\label{incompress1}
& & \nabla\cdot{\bf{v}}=0 ~~~, \\
\label{induction}
& & \frac{\partial{\bf{B}}}{\partial t}=\nabla\times({\bf{v}}\times{\bf{B}}) ~~~, \\
\label{solenoid}
& &\nabla\cdot{\bf{B}}=0 ~~~, 
\end{eqnarray}  
in usual notations. The equations (\ref{stokes})-(\ref{solenoid}) are 
written in dimensionless form, with all variables normalized according to 

\begin{eqnarray}
\label{norm}
& &{\bf{B}}\longrightarrow \frac{{\bf{B}}}{B_0} ~~~,\\
& &{\bf{v}}\longrightarrow \frac{\bf{v}}{v_a}~~~,\\
& & L \longrightarrow \frac{L}{L_0}~~~,\\
& & t \longrightarrow \frac{t}{\tau_a} ~~~,\\
& & p  \longrightarrow \frac{p}{\rho {v_a}^2}~~~. 
\end{eqnarray}
Here, the constants $B_0$ and $L_0$ are generally arbitrary,
but can be fixed by the magnetic field strength and size of the
system. Furthermore, $v_a \equiv B_0/\sqrt{4\pi\rho_0}$ is the Alfv\'{e}n
speed and $\rho_0$ is the constant mass density. The constants $\tau_a$ and
$\tau_\nu$ have dimensions of time, and represent Alfv\'{e}n transit
time ($\tau_a=L_0/v_a$) and viscous diffusion time scale ($\tau_\nu=
L_0^2/\nu$) respectively, with $\nu$ being the kinematic viscosity. 
The pressure $p$ satisfies the elliptic partial differential 
equation
\begin{equation}
\label{pressure}
\nabla^2\left(p+\frac{v^2}{2}\right)=\nabla\cdot\Big[(\nabla\times{\bf{B}})\times{\bf{B}}-(\nabla\times{\bf{v}})\times{\bf{v}}\Big]
\end{equation}
\noindent generated by imposing the incompressibility (\ref{incompress1}) 
on the momentum  transport equation (\ref{stokes}); see \citep{low-bhattacharyya} 
for discussion. If released from an initial non-equilibrium state, the above fluid
evolves by converting  magnetic energy into the energy of mass flow
while the latter gets dissipated by viscosity.  The terminal state is
then expected to be static where the pressure gradient is balanced
by the  Lorentz force as MFLs cannot get diffused because of the
flux-freezing. In simulations however,  only a quasi-steady state is
achieved, maintained by a partial balance of Lorentz force, pressure
gradient and viscous drag; details of the energetics can be found in
\citep{low-bhattacharyya, sanjay-pap, dinesh-2015, complexity-pap}. The
CSs develop as the fluid relaxes to the terminal state. 

As thickness of the developing CSs falls below the selected grid
resolution, scales become under-resolved  and numerical artifacts
such as spurious oscillations are generated through employed numerical
techniques. These under-resolved scales can be removed by utilizing an apt
numerical diffusivity of non-oscillatory finite volume differencing. In
literature, such calculations  relying on the non-oscillatory numerical
diffusivity are referred as  Implicit Large Eddy Simulations (ILESs)
\citep{grinstein-2007}. The computations performed in this work
are in the spirit of ILESs where simulated MRs, being intermittent 
in space and time, mimic physical reconnections in high-$S$ fluids.

The rest of the paper is organized as follows. In section II we construct 
the initial magnetic field and discuss the numerical model. Section III 
is dedicated to results and discussions. In section IV, we summarize 
results and highlight important findings.

\section{Initial magnetic field and numerical model}
\subsection{Initial magnetic field}

\begin{figure}
\centering
\includegraphics[angle=0,scale=.80]{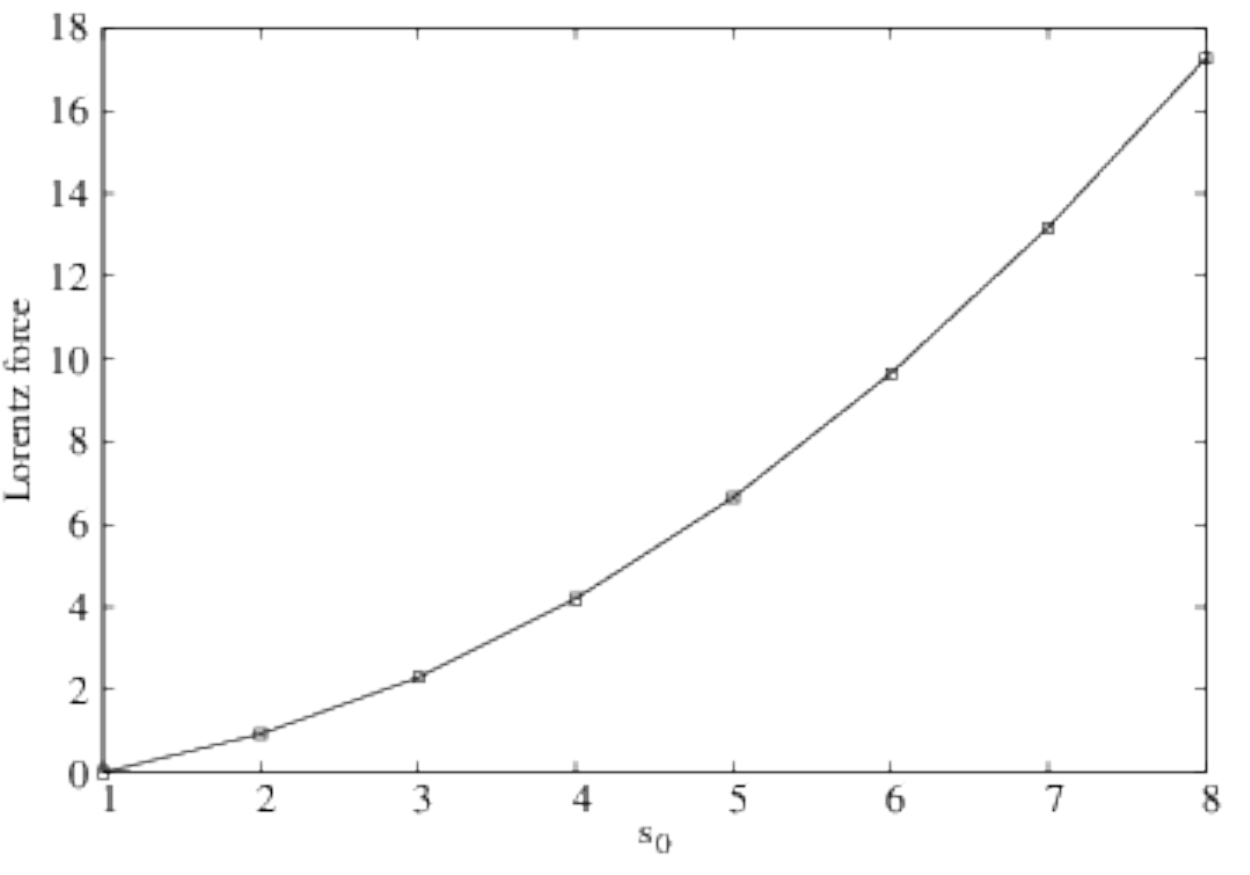}
\caption{Variation of $\mid{\bf{J}}\times{\bf{B}}\mid_{\rm{max}}$ 
with an increase in $s_0$. The plots show a monotonous increase in Lorentz force with an 
increase in $s_0$.} \label{lorentz}
\end{figure}

To construct an initial magnetic field with field-line topology similar to 
coronal loops, we choose ${\bf{B}}({\bf{r}},t=0)$
with the corresponding components
\begin{eqnarray}
\label{comph1}
& & B_x = k_z\sin(k_x x)\exp \left(-\frac{k_z z}{s_0}\right) ~~,\\
\label{comph2}
& & B_y = \sqrt{{k_x}^2-{k_z}^2}\sin(k_x x)\exp \left(-\frac{k_z z}{s_0}\right)~~ , \\
\label{comph3}
& & B_z = s_0k_x\cos(k_x x)\exp \left(-\frac{k_z z}{s_0}\right) ~~, 
\end{eqnarray}
defined in the positive half-space ($z\ge 0$) of a Cartesian domain, 
assumed periodic in $x$. The field is two dimensional as it only 
depends on $x$ and $z$ but not on $y$ and hence, satisfy 
translational symmetry along $y$. The merit of this choice is 
that for $s_0=1$ ${\bf{B}}$ is reduced to a gauge-invariant form 
of the linear force-free field 
\begin{equation}         
\label{lfff}
\nabla\times {\bf{B}_{lf}}=\alpha_0{\bf{B}_{lf}}~
\end{equation}
with the (constant) magnetic circulation per unit flux
$\alpha_0=\sqrt{{k_x}^2-{k_z}^2}$ \citep{parker-ppcf}, and the
associated Lorentz force $(\nabla\times{\bf{B}})\times{\bf{B}}
:={\bf{J}}\times{\bf{B}}\equiv 0$. 
In the conducted simulation, the translational symmetry 
is crucial for an identification of detached magnetic structure with 
flux-rope. It helps to establish the reconnected field lines that generate a magnetic 
flux surface, while providing fundamental understanding 
necessary to explore activation of a three dimensional rope. Standardly, the solar corona is
a low-$\beta$ plasma with Lorenz force dominating all other forces,
and the zero Lorentz force corresponding to a viable equilibrium
\citep{frozen}.  With the initial ${\bf{B}}$ having congruent analytical
form as ${\bf{B}}_{\rm lf}$ (Appendix A), the field lines are expected
to resemble coronal loops. The shear angle between $x$-axis and the
projection of an initial field line on the $z=0$ plane, 
\begin{eqnarray}         
\label{incli2}
& &\phi=\tan^{-1}\left(\frac{\sqrt{{k_x}^2-{k_z}^2}}{k_z}\right)~~,
\end{eqnarray}
is independent of $s_0$. If $k_x=k_z$, $\phi=0$, and the field lines 
are tangential to $y$-constant planes. Generally, the Lorentz force 
exerted by the field defined in (\ref{comph1})-(\ref{comph3}) is 
\begin{eqnarray}
\label{lorentz-nz1}
& &({\bf{J}}\times{\bf{B}})_x =\left[-k_x({k_x}^2-{k_z}^2)+k_x s_0\left(s_0{k_x}^2-\frac{{k_z}^2}{s_0}\right)\right] 
\sin^2(k_x x)\exp\left(-\frac{2k_z z}{s_0}\right)~, \\
\label{lorentz-nz2}
& &({\bf{J}}\times{\bf{B}})_y = 0~, \\
\label{lorentz-nz3}
& &({\bf{J}}\times{\bf{B}})_z = \left[\frac{{k_z}}{s_0}({k_x}^2-{k_z}^2)-k_z\left(s_0{k_x}^2-
\frac{{k_z}^2}{s_0}\right)\right]\frac{\sin(2k_x x)}{2}\exp\left(-\frac{2k_z z}{s_0}\right)~.
\end{eqnarray}
For all $s_0\neq 1$, ${\bf{J}}\times{\bf{B}}\ne 0$, so the Lorentz
force contributes to the viscous relaxation. To assure sheared field lines, we 
set $k_x=1$ and $k_z=0.9$ in the initial field
(\ref{comph1})-(\ref{comph3}). Furthermore, we set $s_0=6$, to optimize
between computation cost and efficient development of dynamics leading
to formation of CSs and their subsequent reconnections. This selection
relied on the monotonous dependence of the maximal initial Lorentz
force  on $s_0$, Figure~{\ref{lorentz}}, and some auxiliary simulations
(not reported here).

\begin{figure}
\centering
\includegraphics[angle=0,scale=.40]{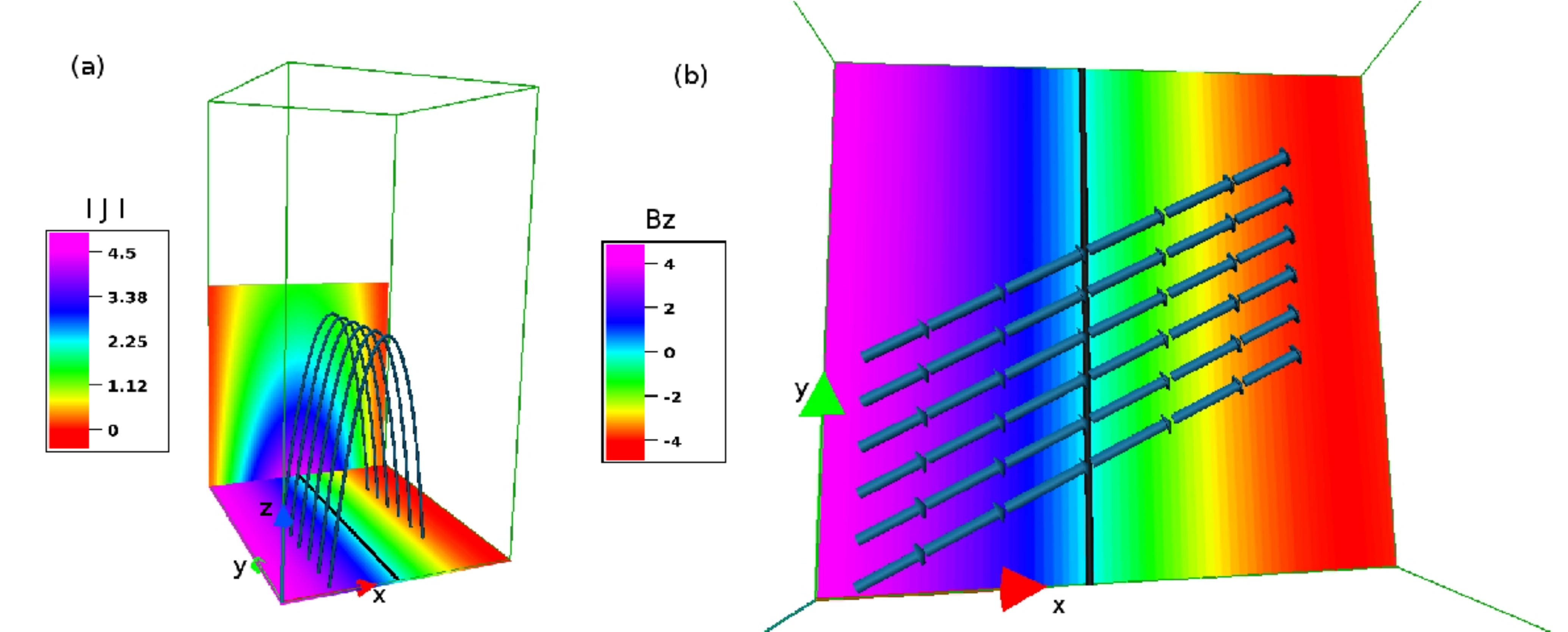}
\caption{The panel a and b illustrate magnetic field lines of the initial field {\bf{B}} for $s_0=6$ and 
their projections of the $z=0$ plane respectively. 
The projected field lines are inclined to the $x$-axis with an angle $\phi$, manifesting the 
sheared nature of MFLs. Both panels are overlaid with contours of the $z$-component of ${\bf{B}}$ ($B_z$)
on the $z=0$ plane along with straight PIL (in color black), indicating polarities of footpoints. 
Additionally, in panel a, contours of magnitude of current density ($\mid {\bf{J}} \mid$) are plotted 
on a $y$-constant plane to depict the current density inside the sheared arcades.  } \label{fields06}
\end{figure}

\begin{figure}
\centering
\includegraphics[angle=0,scale=.40]{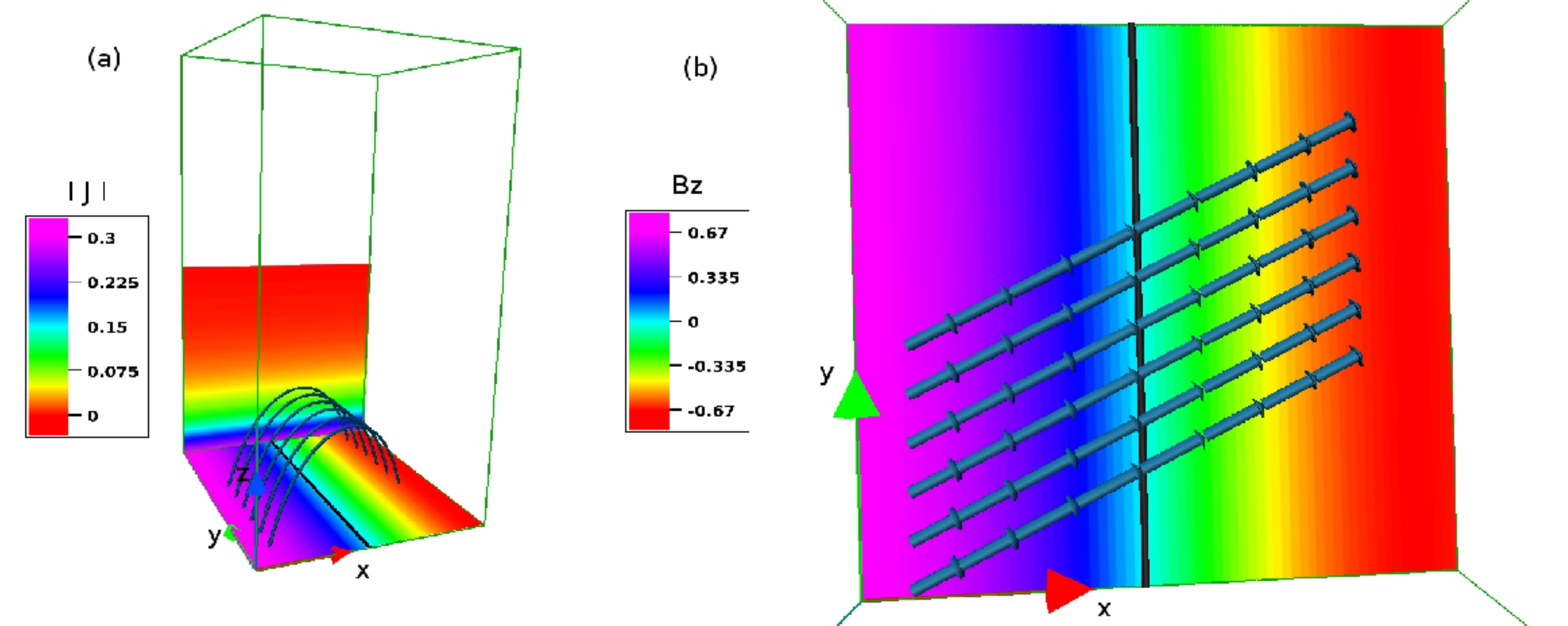}
\caption{As in Fig. \ref{fields06} but for field lines of ${\bf{B}}_{\rm lf}$ and their projections
 on the $z=0$ plane. The figure demonstrates the morphology 
and shear of the initial field ${\bf{B}}$ to be identical to the corresponding   
${\bf{B}}_{\rm lf}$  characterized by $s_0=1$. } \label{fields01}
\end{figure}

To aid further understanding, the following presents a detailed
analysis of the initial magnetic field.  For relevant depictions,
hereafter, we set $x\in\{0,\pi\}$, because similar structures  and
dynamics are repeated in $x \in \{\pi, 2\pi\}$ due to the assumed
periodicity. The MFLs for $s_0=6$ are shown in Figure {\ref{fields06}}
(a).\footnote{The VAPOR visualization package {\citep{clyne-rast}}
is used to integrate the field lines equations.} The arrows in 
colors red, green, and blue denote the axes $x$,
$y$, and $z$ respectively. The plotted MFLs are sheared loops with  a
straight polarity inversion line (PIL) located at $(x,z)=(\pi/2,0)$.  To
highlight the shear, Figure {\ref{fields06}} (b) illustrates projection
of the  field lines on $z=0$ plane which are inclined to the $x$-axis
at an angle  $\phi\equiv\tan^{-1}(0.48)=25.6^{\circ}$. We keep $\phi$
fixed to this value for our simulations. Notably, the MFLs maintain
translational symmetry along $y$ since ${\bf{B}}$ is  independent of
$y$.  For comparison, in panels a and b of Figure {\ref{fields01}},
we plot MFLs of the corresponding ${\bf{B}}_{\rm lf}$ (for $s_0=1$)
and their projections on the $z=0$ plane.

Coronal arcades---associated with flux rope formation---are generally 
believed to evolve from an initial quasi-equilibrium state,
devoid of any major electric current density. 
An appropriate  physical mechanism is then required to generate dynamics from the  
quasi-equilibrium. Such onset of dynamics can be achieved 
by a specialized photospheric flow used in \citep{devore2000},
where fully compressive MHD simulations demonstrate generation of ropes through reconnections. 
The present computations, however, start from an 
initial non-equilibrium state with appreciable electric current density to make them harmonious with 
general framework of the viscous relaxation.
For instance,  the initial electric 
current density is roughly 15 times larger for the ${\bf{B}}$ in comparison to the ${\bf{B}}_{\rm lf}$ 
(cf. Figs. \ref{fields06} and \ref{fields01}).

\subsection{Numerical model}

\begin{figure}
\centering
\includegraphics[angle=0,scale=.60]{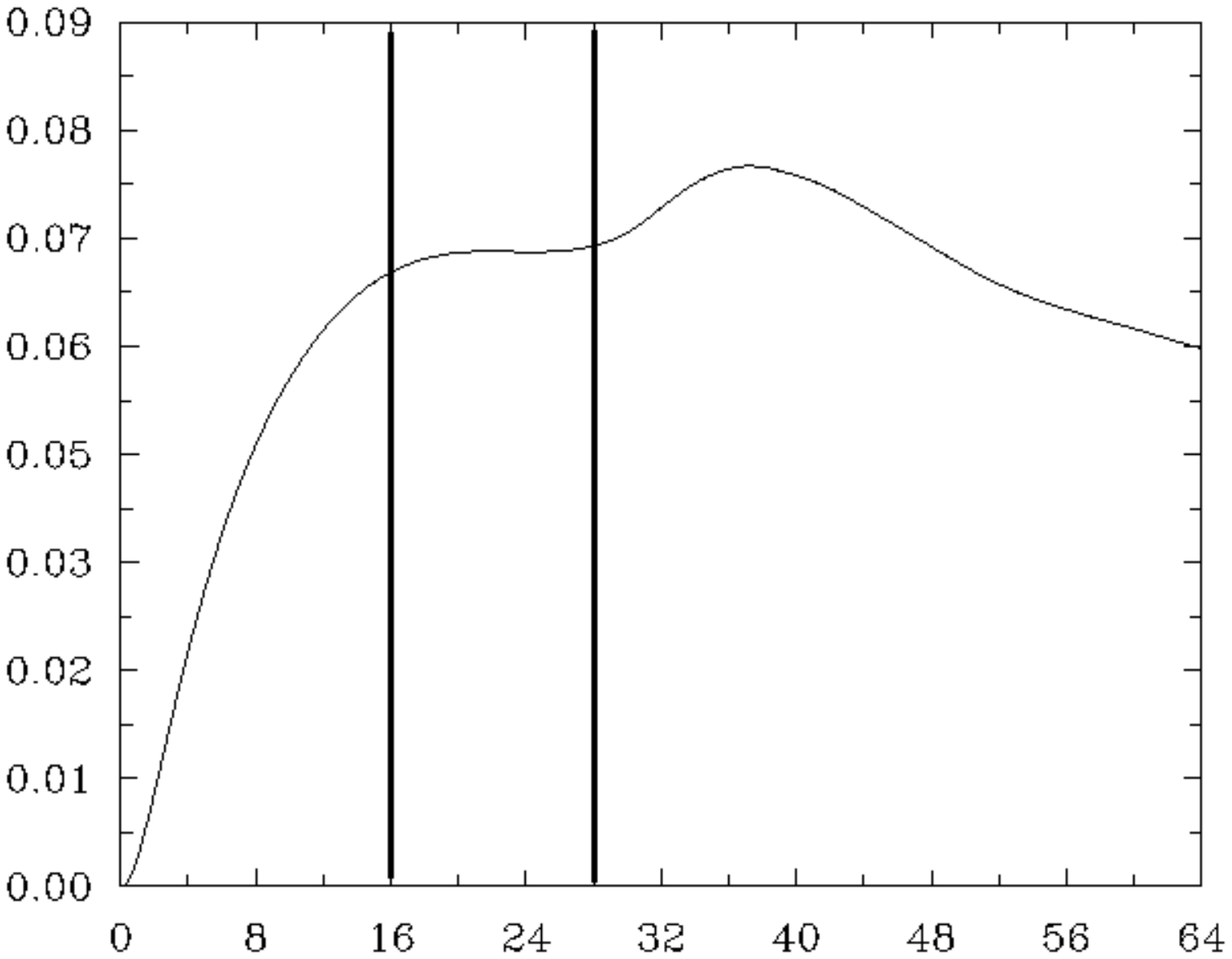}
\caption{ The evolution of kinetic energy, normalized to initial total 
(kinetic+magnetic) energy. Noteworthy are the three distinct phases of 
the evolution, marked by vertical lines in the plot. } \label{kinetic}
\end{figure}

\begin{figure}
\centering
\includegraphics[angle=0,scale=.30]{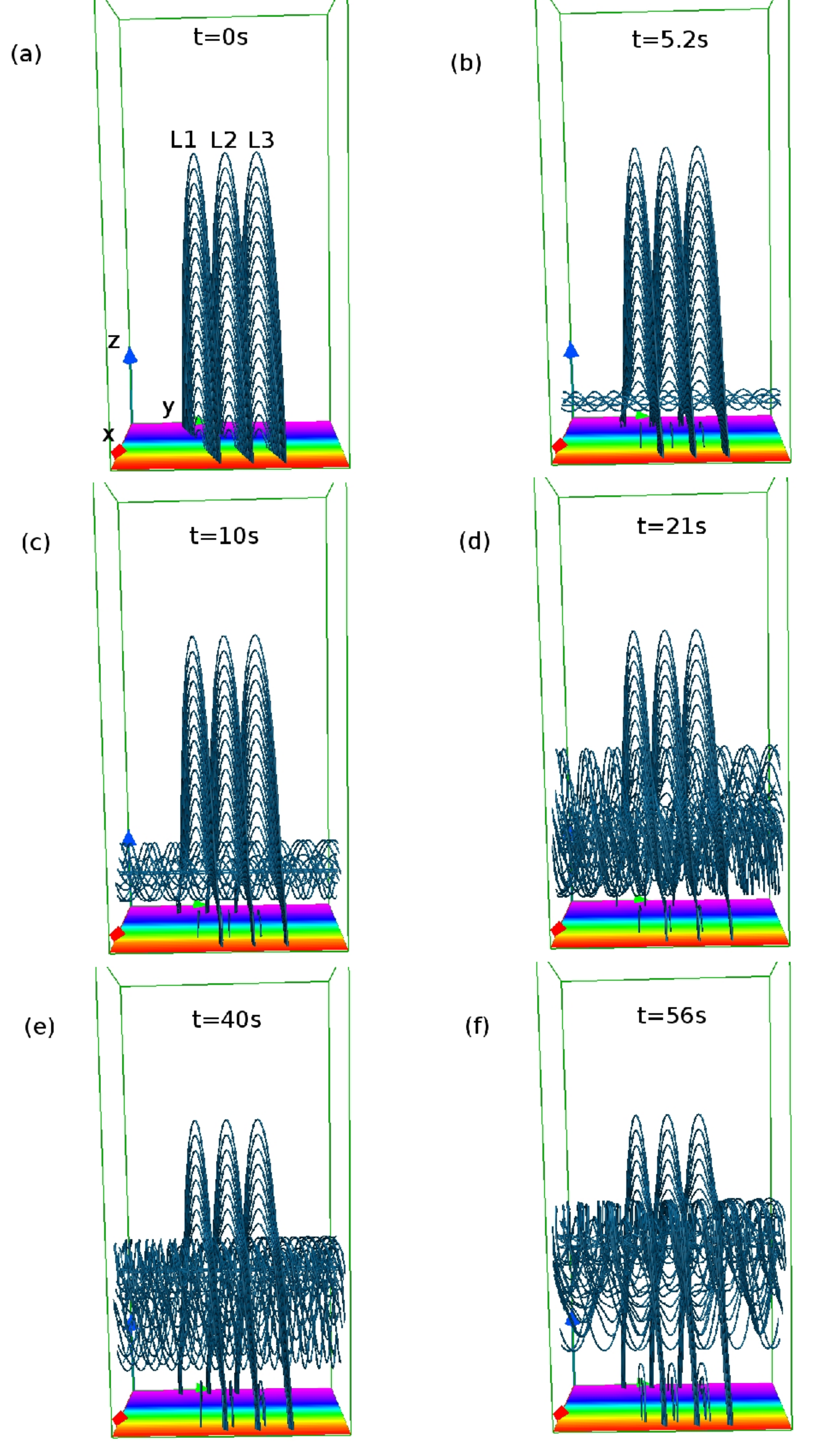}
\caption{ Time sequences of  magnetic field lines in their important phases of evolution. 
Three different sets of magnetic loops are sketched, denoted by $L_1$, $L_2$ and $L_3$ in panel a of the
figure. Important is formation of detached magnetic structure resembling a magnetic flux-rope.
Also evident is the ascend of this structure while being situated over the PIL. } \label{mfls}
\end{figure}

\begin{figure}
\centering
\includegraphics[angle=0,scale=.30]{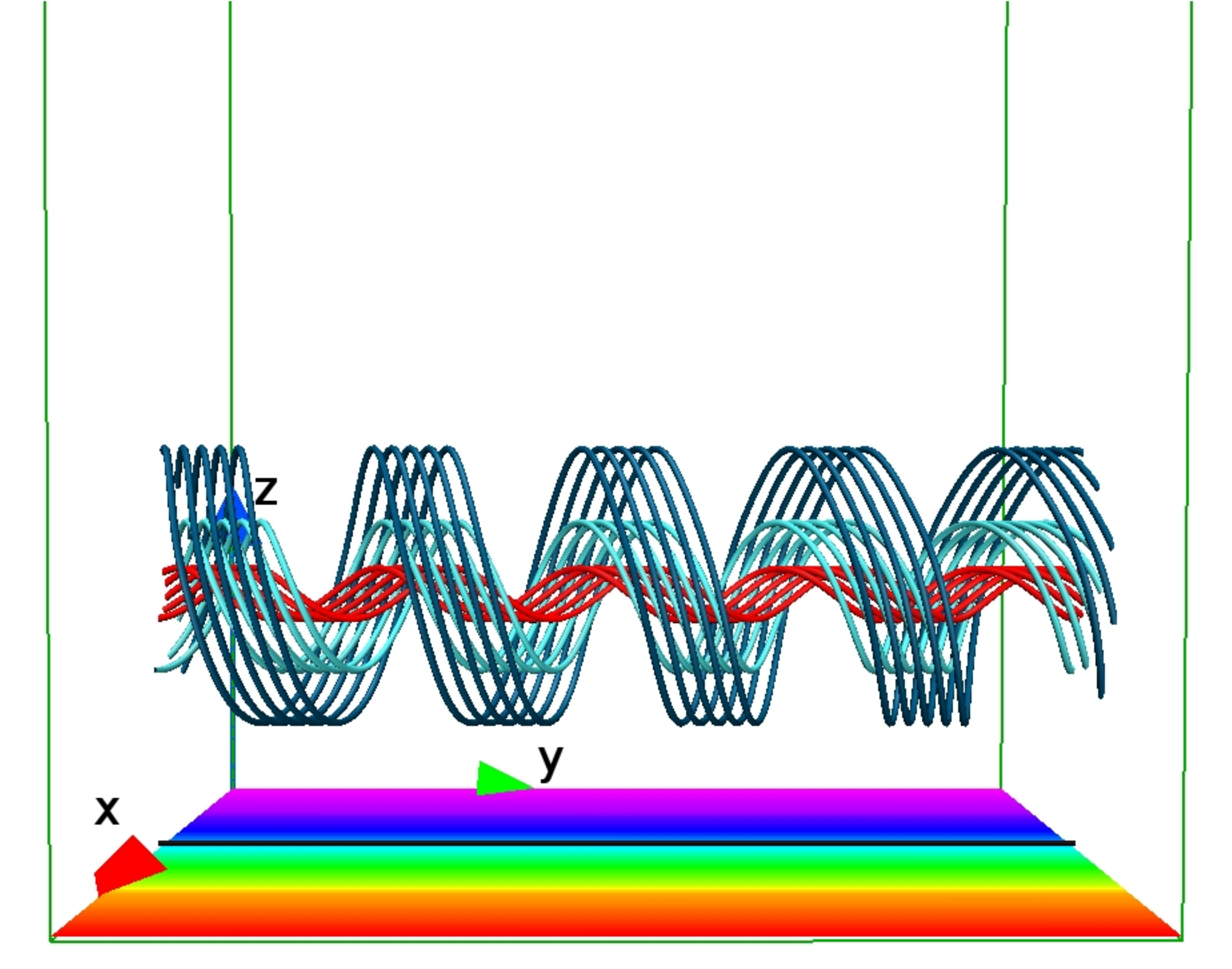}
\caption{Snapshot of field lines at $t=10s$, plotted in the neighborhood of the detached structure.
The figure verifies the detached magnetic structure to be comprised of a stack of co-axial 
cylindrical magnetic flux surfaces which are made of helical field lines. } \label{flux-rope}
\end{figure}

\begin{figure}
\centering
\includegraphics[angle=0,scale=.40]{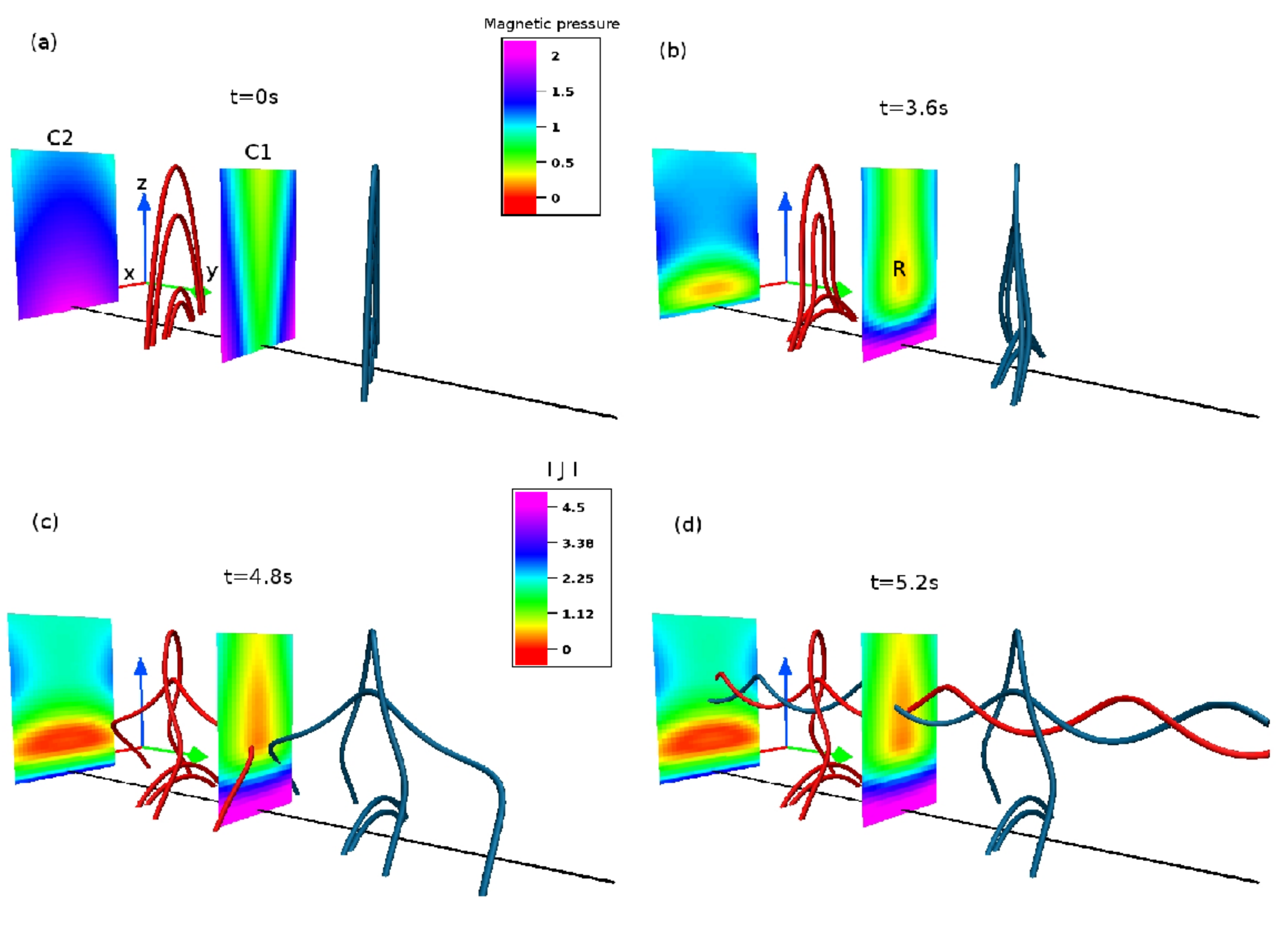}
\caption{Evolution of magnetic field lines concurrent with the first
phase. The figure is overlaid  with contours of magnetic pressure (marked by C1 in panel a) 
and magnitude of current density (marked by C2 in panel a) drawn 
on different $y$-constant planes.  The  PIL is the solid black line.   Panel b
documents the implosion of MFLs situated at lower  height along with
an increase in their footpoint separation which results in depletion of
magnetic pressure (symbolized by R) in the $y$-constant plane. Parts of MFLs
are dragged into this pressure depleted region R from both sides  of PIL,
which sharpens up the gradient in ${\bf{B}}$ (panel c).  Subsequent
reconnection leads to the generation of detached helical field lines
(panel d). The current contours confirm the absence of an extended CS.} \label{initialrec}
\end{figure}

To solve the Navier-Stokes  MHD equations (\ref{stokes})-(\ref{solenoid}),
we  utilize the well established magnetohydrodynamic numerical
model EULAG-MHD {\citep{smolar-2013}}, an extension of the
hydrodynamic model EULAG predominantly used in atmospheric
and climate research \citep{smolar-2006,prusa-2008}. Here we
summarize only essential features of EULAG-MHD; the details are in
Ref.~{\citep{smolar-2013}} and references therein. The EULAG-MHD is
based on the spatio-temporally second order accurate non-oscillatory
forward-in-time multidimensional positive definite advection 
transport algorithm, MPDATA, {{\citep{smolar-2006}}. Relevant to our
simulations is MPDATA proven dissipative property, intermittent and
adaptive to generation of under-resolved scales in field variables for
a fixed grid resolution. With fixed grid resolution, a development of
CSs inevitably generates  under-resolved scales as a consequence of
unbounded increase in the magnetic field gradient. The MPDATA then
removes these under-resolved scales by producing locally effective
residual dissipation of the second order, sufficient to sustain
monotonic nature of the solution. Being intermittent and adaptive,
the residual dissipation facilitate the model to perform ILESs that
mimics the action of explicit subgrid-scale turbulence models, whenever
the concerned advective field is under-resolved \citep{margolin}. Such
ILESs performed with the model have already been successfully utilized
to simulate regular solar cycles {\citep{ghizaru}}, with the rotational
torsional oscillations subsequently characterized and analyzed in 
{\citep{Beaudoin}}. The simulations reported continue relying on the 
effectiveness of ILES in regularizing the onset of MRs, concurrent 
and collocated with developing CSs {\citep{kumar-bhattacharyya}}. 
 
\section{Simulation results}

The simulations are performed on four differently sized grids---$128\times
128\times 256$, $64\times 64\times 128$, $40\times 40\times 80$
and $32\times 32\times 64$---resolving the computational domain
${0,2\pi}\times {0,2\pi}\times {0,8\pi}$ (respectively in $x$, $y$,
and $z$), all starting from a motionless state and initial field
(\ref{comph1})-(\ref{comph3}). The boundary conditions along $x$ are
periodic, and open in $z$. The $y$ direction is analytically ignorable,
because the governing equations (\ref{stokes})-(\ref{solenoid}) and the
initial conditions assure all dependent variables invariant in $y$.
However, for efficacy of the results postprocessing and their analysis
with the VAPOR visualization package {\citep{clyne-rast}} we allow
field variables to have all the three components while circumventing
discrete differentiations in $y$. This makes field variables to have a
translational symmetry along $y$, the maintenance of which can easily be
verified from relevant plots presented in the paper. Moreover, for the
case $32\times 32\times 64$ we have performed a fully 3D simulation,
with no ignorable coordinate, and found results (not shown) to be in
exact agreement with the corresponding 2.5D run. 

In the conducted simulations, the dimensionless number $\tau_a/\tau_\nu
\approx 10^{-5}$, and the effective $S^{-1}$ is negligibly  small apart
from reconnections. The residual dissipation being  intermittent in
time and space, a quantification of it is  meaningful only in the
spectral space where, in analogy to the eddy-viscosity of explicit
subgrid-scale models for turbulent flows, it only acts on the  shortest
modes  admissible on the grid \citep{domaradzki};  in particular,
in the vicinity of steep gradients in simulated fields. For parameter
values relevant to solar corona (Aschwanden 2004),  the ratio of
Alfv\'{e}n transit time to the viscous time scale $\tau_a/\tau_v\approx
10^{-4}$, which is one order of magnitude larger than that adopted in
our computations. This, however, only affects the interval between
two successive reconnections without  an effect on the corresponding
change in field line topology. Being incompressible, flow in our
computations is volume preserving---an assumption also used in other
works \citep{dahlburg, aulanier}. While compressibility is important
for the thermodynamics of coronal loops \citep{ruderman}, our focus is
on elucidating changes in magnetic topology idealised with a thermally
homogeneous magnetofluid. We select $\rho_0=1$ to have $\tau_a \approx
20s$, roughly corresponding to the coronal value with  constants $B_0$
and $L_0$ set to 4 (amplitude of ${\bf{B}}$) and $8\pi$ (vertical length
of the physical domain) respectively. The results for computation with
resolutions  $128\times 128\times 256$ are presented below. 
 
To obtain overall understanding of the evolution, in Figure
{\ref{kinetic}},  we plot the time profile of normalized kinetic
energy. The time profile shows four distinct phases which can  
approximately be divided into three overlapping intervals ranging from
$t \in \{0s,16s\}$, $t \in \{16s,28s\}$, and $t>28s$; separated by the
vertical lines in the figure.   The first phase corresponds to a rise
in kinetic energy as the initial Lorentz force pushes the magnetofluid
from rest. This rise is then arrested by viscous drag and the fluid
settles down to a quasi-steady state  which is characterized by an
almost constant kinetic energy, representing the onset of the second
phase.  Also to be noted are the further rise and the following decay
of kinetic energy which correspond  to the third  phase.

Figure {\ref{mfls}} illustrates three sets of magnetic loops L1, L2,  and L3, the evolution 
of which leads to a topologically distinct structure with field lines detached 
from the $z=0$ surface. This detached structure resembles a magnetic flux-rope 
propagating along the $y$. For substantiation, in 
Figure {\ref{flux-rope}} we plot the field lines in vicinity of the detached structure.
The plot establishes the detached structure to be a flux-rope, made by helical
field lines that are tangential to nested co-axial cylindrical 
surfaces. Away from the axis, the helices are more tightly wound. 
Importantly, the Figure {\ref{mfls}} documents a sustained ascent of
the flux-rope along the  vertical while being always situated above the
PIL. The latter is a general requirement   for magnetic structures to
represent solar prominences or filaments. 

Toward an explanation for generation of the rope, notable is the
implosion  of MFLs with a simultaneous increase in their footpoint
(the point at which MFLs intersect the $z=0$ plane) separation as
displayed in panel b of Figure \ref{initialrec}.  During the implosion,
reconnection of footpoints is prohibited because only parallel MFLs
are bundled together.  With magnitude of the Lorentz force diminishing
exponentially along the vertical, the implosion is non-uniform,  being
more effective at lower heights. The non-uniform implosion generates a
void depleted of MFLs, leading to a local decrease in magnetic pressure 
on a $y$ constant plane; cf. panel b of Figure \ref{initialrec}.
Consequently, parts of two complementary anti-parallel field lines,
located on opposite sides of the PIL, are stretched along $x$ and
enter into this void (panel c). As a result, the gradient of {\bf{B}}
along $x$ sharpens up, as confirmed by the Figure {\ref{scaling}} that
documents  scaling of current density with resolution in the vicinity
of the void. Consequently, magnetic reconnection takes place as the
scales become under-resolved (panel d of Fig. \ref{initialrec}). 
Such reconnections, repeated in time, are responsible for the origin
of  the rope. Crucial is the non-zero shear, or equivalently $B_y\neq  0$,
which makes the  reconnected field lines helical. In absence of shear,
MFLs would have been closed disjoint curves tangential to $y$-constant
planes---thus corresponding to a flux tube, but not a rope.

The projection of the rope on a $y$-constant plane corresponds to a
magnetic island. In Figure \ref{xtype} we plot evolution of the island at
$t=6s$ and $t=8s$.  Important is the appearance of an $X$-type magnetic
null  located below the island; illustrated in Figure {\ref{xtype}} (a)
by the symbol $X$. The MRs at the $X$-type neutral point increase number
of MFLs constituting the island (panel b of Fig. {\ref{xtype}}). Also,
the outflow generated by these repeated MRs at  the $X$-point lifts the
island center along the vertical. Notably, the above MRs are occurring
while preserving the $X$-type null and no extended CS is developed, 
as documented by contours of current density in the Figure \ref{initialrec}.  
The development of an extended CS by squashing a $X$-type null requires
a favorable force missing in this period of evolution.

In Figure {\ref{extndcs}}, we illustrate MFLs projected on a $y$-constant
plane (panels a and b) and contours of current density (panels c and d) at 
instances $t=16s$ and $t=21s$, which correspond to the second phase of the 
evolution. From the figure, as  reconnections at
the $X$-type null continue, the  magnetic pressure, below and above
the two quadrants of the $X$-type null (denoted by $X1$ and $X2$ in
panel a of the figure), increases. The increased magnetic pressure
results in squashing of the $X$-type null and  leads to formation of two
$Y$-type nulls along with an extended CS (panels b and d).  Importantly, the
development of this extended CS is concurrent with the quasi-steady phase
of the evolution $t \in \{16s,28s\}$, as demanded by the magnetostatic
theorem. The decay of this CS  is responsible for the post quasi-steady
rise in kinetic energy, marking onset of the  third phase.

The corresponding dynamics of  MFLs in the third phase is  documented
in Figure {\ref{ascend}} for time instances $t=30s$, $t=40s$, $t=50s$,
and $t=55s$. Noticeably, the reconnection at the extended CS reduces
the  pressure beneath the flux-rope because of a localized decrease in
magnetic field strength.  The neighboring field lines are stretched
into this pressure depleted region, from all sides, rendering the
rope to be dipped at the bottom portion.  This dip corresponds to
the observed dipped portion of a  prominence, where the mass of the
prominence is believed to be situated {\citep{van&cranmer}}. Along with
the stretched field lines located below the pressure depleted region, the
dipped portion of the rope  generates a new $X$-type null---denoted by
X3 in the figure. Along with the rope, this new $X$-type null also moves
upward. Further evolved, the MFLs constituting the flux-rope reconnect
and the rope loses its well defined structure.  

Noteworthy is importance of the translational symmetry in identifying the detached magnetic structure 
with a flux-rope---which is first and foremost a magnetic flux surface by its mathematical definition. 
Toward recognizing the importance, projection of a helical field line constituting the 
detached structure (Figure \ref{flux-rope}) on a $y$-constant 
surface is a closed curve. Because of the symmetry, a translation of this closed curve along 
$y$ generates a surface on which the helical field lines are also tangential; confirming the detached structure to be
a magnetic flux surface.   
In absence of the symmetry, the identification is not straightforward as field lines are 
always postprocessed and the processing error contributes to the topology of the obtained MFLs.
Conventionally, however, the magnetostatic theorem applies to three dimensional fields, which 
favors the actual evolution of solar magnetized plasma.
To consolidate the simulation results further, 
in the following we present two auxiliary computations  where the symmetry is removed. 
The  three dimensional simulations are performed on a grid of size $128\times
128\times 256$, resolving the same computational domain as in the symmetric case. 
The boundary conditions along $x$ and $y$ are periodic, and open in $z$. For consistency, all other parameters are kept identical
to the symmetric case.

\subsection{Auxiliary simulation I}
For the first simulation, the initial magnetic field ${\bf{B}}^\star$ (Appendix B) is constructed by superposing a three dimensional 
solenoidal field on the ${\bf{B}}$ expressed in equations (10)-(12). 
Notable is the structural similarity of the two superposed fields that 
individually reduce to a linear force free field for $s_0=1$. 
The evolution for this 3D simulation is shown in Figures \ref{3dinitialrec} and  \ref{3dmov}.
Figure \ref{3dinitialrec} displays the time series for two sets of field lines, where panel a corresponds 
to the initial field ${\bf{B}}^\star$. The PIL is curved, thus attesting to the absence of translational symmetry along the $y$.  
The figure is overlaid with contours of magnetic pressure drawn on a $y$-constant plane. 
Similar to evolution depicted in Figure 7, repeated reconnections 
generate a detached magnetic structure which, based on the 2.5D computation, can be identified to a flux-rope. 
The Figure \ref{3dmov} illustrates the evolution with more densely plotted field lines where panel a corresponds to the initial field 
${\bf{B}}^\star$ (corresponding animation is provided as supplementary material). The field lines constituting the rope are marked in red, and 
with an ascent maintained by underlying reconnections document an evolution with overall similarity to its 2.5D counterpart.

\subsection{Auxiliary simulation II}
For the second simulation, the initial field ${\bf{B}}^{\star\star}$ (Appendix C) is derived by 
superposing the ${\bf{B}}^\star$ with another solenoidal field ${\bf{B}}^\prime_p$ where the ${\bf{B}}^\prime_p$ 
reduces to potential field for $s_0=1$.  
In Figure \ref{fluxrope3} we illustrate the evolution of magnetic field lines (corresponding animation is provided as 
supplementary material). 
Panel a depicts the initial field lines of ${\bf{B}}^{\star\star}$ having a curved PIL. 
The figure confirms the formation of flux rope (marked in red) and its 
ascent by repetitive reconnections similar to the first 3D simulation (Figure \ref{3dmov}). 
The figure is overplotted with contours of $\mid{\bf{B}}^{\star\star}\mid$ in $y$-constant plane and isosurfaces of 
corresponding current density ${\bf{J}}^{\star\star}$ with isovalues $15\%$ and $20\%$ of maximum $\mid{\bf{J}}^{\star\star}\mid$.
Based on their appearances,  the isosurfaces can be classified into two distinct categories: the surfaces
appearing at the $z$-constant plane below the rope (marked by A in panel d) and the elongated surfaces located at 
horizontal sides of the rope (marked by B in panel d). The figure identifies generation of the elongated surfaces to an increase 
in the local number density of parallel field lines, resulting in an increase in $\mid{\bf{B}}^{\star\star}\mid$ and hence $\mid{\bf{J}}^{\star\star}\mid$ without 
any sharpening  of field gradient. Hence, these surfaces do not represent current sheets \citep{complexity-pap}.  
In contrast, the surfaces lying in $z$-constant plane originate without a co-located 
enhancement in $\mid{\bf{B}}^{\star\star}\mid$, suggesting the development of these surfaces by 
a local increase in the field gradient. Thus, the appearances of these surfaces indicate the 
formation of current sheets below the rope \citep{complexity-pap}. Importantly, the current sheet has a non-uniform 
intensity distribution along the rope, where patches of intense currents are marked in black while low currents are marked in yellow; which agrees with the
general expectation. The same non-uniformity of developing CSs was also observed in the Auxiliary simulation I (not shown) which further validates 
the general expectation.

With the ascending flux ropes having underlying reconnections in agreement with the standard flare
model \citep{shibata2011}, the reported simulations identify spontaneous repeated
MRs as the initial driver for  the rope formation and triggering its
ascent. Further, the simulations underline that MRs play an active role
in the feedback mechanism between flux-rope dynamics and reconnections,
central to the standard flare model. This is in harmony with contemporary
observations \citep{temmer2008, temmer2010, cho2009, cheng2011}}.

\section{Summary and conclusions}

The presented simulations start from a select motionless state with
magnetic field congruent to a gauge-invariant form of linear force-free
field having translational symmetry. 
These simulations identify repeated magnetic reconnections as an
autonomous mechanism for creating a flux-rope from initially bipolar
field lines and, subsequently, for triggering and maintaining its ascent
via reconnections that occur below the rope. The computations being
commensurate with the requirements of magnetostatic theorem, the
reconnections are spontaneous and inherent to the evolving fluid. The
computational commensuration with the  Parker's theorem is achieved by
viscous relaxation of an incompressible, thermally homogeneous high-$S$
magnetofluid maintaining the condition of flux-freezing. During
the relaxation, sharpening of magnetic field gradient is unbounded,
ultimately leading to MRs at locations where separation of 
non-parallel field lines approaches grid resolution. The MR process per
se is underresolved,  but effectively regularized by locally adaptive
dissipation of MPDATA,  in the spirit of ILES subgrid-scale turbulence
models. In effect,  the post-reconnection condition of flux-freezing
is restored, and field lines tied to the reconnection outflow push
other sets of MFLs,  leading to secondary MRs. The whole process is
replicated in time to  realize repetitive reconnections.

The simulated relaxation process comprises three distinct phases. In the
first phase, a combination of incompressibility and the initial Lorentz
force deforms initial field lines such that the field gradient sharpens
in a direction implied by the initial condition (herein $x$). Further
push  eventuates in reconnection and development of a $X$-type neutral
point along  with a detached flux-rope. Repeated reconnections around
the $X$-type  null generate more detached field lines which contribute
to the rope.  Moreover, being frozen to the outflow, the rope ascends
vertically.  Because the reconnections are localized below the evolving
rope,  the scenario is in general agreement with observations. As the
magnetofluid relaxes to a quasi-steady state, the process enters the 
second phase of the relaxation with the $X$-type null being squashed to
generate two $Y$-type nulls along with an extended CS beneath the rope.
In the third phase the extended CS decay, resulting in an increase in the
kinetic energy of mass flow. Because of the corresponding decrease in
magnetic intensity near the decaying CS, MFLs from all side of the CS
are stretched into the field depleted region. The rope becomes dipped at
the bottom and a new $X$-type null is generated which ascends with the
rope. Continued  further in time, the flux-rope reconnects internally
and loses its structure.  Fully three dimensional simulations are
 also performed to verify the robustness of repeated spontaneous MRs 
in creation and ascent of a rope accordant to a more realistic evolution. 
Importantly, the three dimensional simulations document the intensity 
of CSs developed over respective PILs to be non-uniform---a feature 
expected in realistic ropes developed via reconnections. 

Altogether the computations extend Parker's magnetostatic theorem to
the scenario of evolving magnetic fields which can undergo magnetic
reconnection. Notably, the theorem in absence of magnetic diffusivity
leads to CSs, having true mathematical singularities in magnetic
field which are end states of any evolution. But in presence of
small but non-zero magnetic diffusivity, as in astrophysical plasmas,
the theorem opens up the possibility of spontaneous generation of
secondary CSs and subsequent MRs that may contribute  to the dynamics
of the plasma. The computations confirm the contribution
to be meaningful as it can generate observed magnetic structures and
govern their dynamics which, in this case, is the evolution of a
flux-rope in terms of its generation and ascent in a magnetic topology
relevant to the solar corona. Additionally, in context of the standard
flare model, the simulations imply a direct involvement of magnetic
reconnection in the activation of flux-ropes. 

\acknowledgements 

\noindent {\bf Acknowledgements:} The simulations are performed using the 100 TF
cluster Vikram-100 at Physical Research Laboratory, India. One of us (PKS) is supported by
funding received from the European Research  Council under the European
Union's Seventh  Framework Programme (FP7/2012/ERC Grant agreement
no. 320375). The authors also sincerely thank an
anonymous reviewer for providing specific suggestions to
enhance the presentation as well as to raise the academic
content of the paper.

\appendix

\section{Appendix}

To solve a linear force-free equation, the magnetic field is
identified as a Chandrasekhar-Kendall \citep{chandra} eigenfunction
which, in Cartesian coordinate can be written as
\begin{equation}
{\bf{Y}}=\nabla \times \psi\hat{e_y}+\frac{1}{\alpha_0}\nabla\times(\nabla\times \psi\hat {e_y}) ~~,
\end{equation}
where $\psi=\psi(x,y,z)$ is a scalar function and $\alpha_0$ is constant. 
Substitution of ${\bf{Y}}$ in the linear force-free equation gives,
\begin{equation}
\nabla \times (\nabla\times\nabla\times\psi\hat{e_y}-{\alpha_0}^2\psi\hat{e_y})=0 ~~ ,
\end{equation}
implying,
\begin{equation}
(\nabla\times\nabla\times\psi\hat{e_y}-{\alpha_0}^2\psi\hat{e_y})=\nabla\lambda ~~,
\end{equation}
where the scalar function $\lambda=\lambda(x,y,z)$ represents an arbitrary gauge.
Using the identity 
\begin{equation}
\nabla\times\nabla\times{\psi\hat{e_y}}=-\nabla^2{\psi\hat{e_y}}+\nabla(\nabla\cdot{\psi\hat{e_y}})~~,
\end{equation}
results in the inhomogeneous vector Helmholtz equation for the magnetic field
\begin{equation}
({\nabla}^2\psi+{\alpha_0}^2\psi)\hat{e_y}= \nabla(\nabla\cdot \psi \hat{e_y}-\lambda)~~. 
\end{equation}

Standardly, a three dimensional analytical solution of the linear force-free 
equation is obtained by only solving the homogeneous equation which corresponds 
to the selection of 
\begin{equation}
\label{gauge}
\lambda= \nabla\cdot \psi \hat{e_y}~,
\end{equation}
for the gauge. If $\partial /\partial y \equiv 0$, the solution for the linear force-free 
equation is gauge independent, and the scalar function $\psi$ satisfies Helmholtz equation 
\begin{equation}
\nabla^2\psi+\alpha_0^2 \psi =0~~~. 
\end{equation}
Consequently, for a geometry relevant to solar corona, the field components are 
\begin{eqnarray}
\label{lcomph1}
& & {B_{\rm lf}}_x = k_z\sin(k_x x)\exp \left(-k_z z\right) ,\\
\label{lcomph2}
& & {B_{\rm lf}}_y = \sqrt{{k_x}^2-{k_z}^2}\sin(k_x x)\exp \left(-k_z z\right) , \\
\label{lcomph3}
& & {B_{\rm lf}}_z = k_x\cos(k_x x)\exp \left(-k_z z\right),
\end{eqnarray}
with $\alpha_0=\sqrt{{k_x}^2-{k_z}^2}$.

\section{Appendix}
A three dimensional linear force-free field ${\bf{B}}^{\prime}_{lf}$, with the choice of gauge, has components

\begin{eqnarray}
\label{hlcomph1}
& & {B^{\prime}_{lf}}_x = \alpha_0 l_y\sin(l_x x)\cos(l_y y)\exp \left(-l_z z\right)-
                       l_x l_z\cos(l_x x)\sin(l_y y)\exp \left(-l_z z\right)  ,\\
\label{hlcomph2}
& & {B^{\prime}_{lf}}_y = -\alpha_0 l_x\cos(l_x x)\sin(l_y y)\exp \left(-l_z z\right)-
                       l_y l_z\sin(l_x x)\cos(l_y y)\exp \left(-l_z z\right)  ,\\
\label{hlcomph3}
& & {B^{\prime}_{lf}}_z = ({l_x}^2+{l_y}^2)\sin(l_x x)\sin(l_y y)\exp \left(-l_z z\right),
\end{eqnarray}
with $\alpha_0=\sqrt{{l_x}^2+{l_y}^2-{l_z}^2}$. The three dimensional simulation is performed with an initial field 

\begin{equation}
\label{initial2}
{\bf{B}}^\star={\bf{B}}+a_0{\bf{B}}^{\prime}
\end{equation}
for $a_0=.5$ and ${\bf{B}}^{\prime}$ having components 

\begin{eqnarray}
\label{3dlcomph1}
& & {B}^{\prime}_x = \sin(x)\cos(y)\exp \left(-\frac{z}{s_0}\right)-
                      \cos(x)\sin(y)\exp \left(-\frac{z}{s_0}\right)  ,\\
\label{3dlcomph2}
& & {B}^{\prime}_y = -\cos(x)\sin(y)\exp \left(-\frac{z}{s_0}\right)-
                       \sin(x)\cos(y)\exp \left(-\frac{z}{s_0}\right)  ,\\
\label{3dlcomph3}
& & {B}^{\prime}_z = 2s_0\sin(x)\sin(y)\exp \left(-\frac{z}{s_0}\right) .
\end{eqnarray}

\section{Appendix}
The force-free field ${\bf{B}}_{\rm lf}$ (Appendix A) reduces to a two-dimensional potential field 
${\bf{B}}_p$ with $k_x=k_z=k$. Then, the components of the ${\bf{B}}_p$ are     

\begin{eqnarray}
\label{pcomph1}
& & {B_{p}}_x = k\sin(k x)\exp \left(-k z\right) ,\\
\label{pcomph2}
& & {B_{p}}_y = 0 , \\
\label{pcomph3}
& & {B_{p}}_z = k\cos(k x)\exp \left(-k z\right),
\end{eqnarray}

\noindent  with $\alpha_0=0$. The additional three dimensional simulation is conducted with the initial field  

\begin{equation}
\label{initial3}
{\bf{B}}^{\star\star}={\bf{B}}^\star+{\bf{B}}_p^{\prime}
\end{equation}

\noindent where ${\bf{B}}_p^{\prime}$ (derived from the ${\bf{B}}_p$) has the components

\begin{eqnarray}
\label{npcomph1}
& & {B_{p}}^\prime_x = \sin(x)\exp \left(-\frac{z}{s_0}\right) ,\\
\label{npcomph2}
& & {B_{p}}^\prime_y = 0 , \\
\label{npcomph3}
& & {B_{p}}^\prime_z = s_0\cos(x)\exp \left(-\frac{z}{s_0}\right).
\end{eqnarray}

\clearpage

\clearpage

\clearpage

\clearpage

\clearpage

\clearpage

\clearpage

\clearpage

\begin{figure}
\centering
\includegraphics[angle=0,scale=.80]{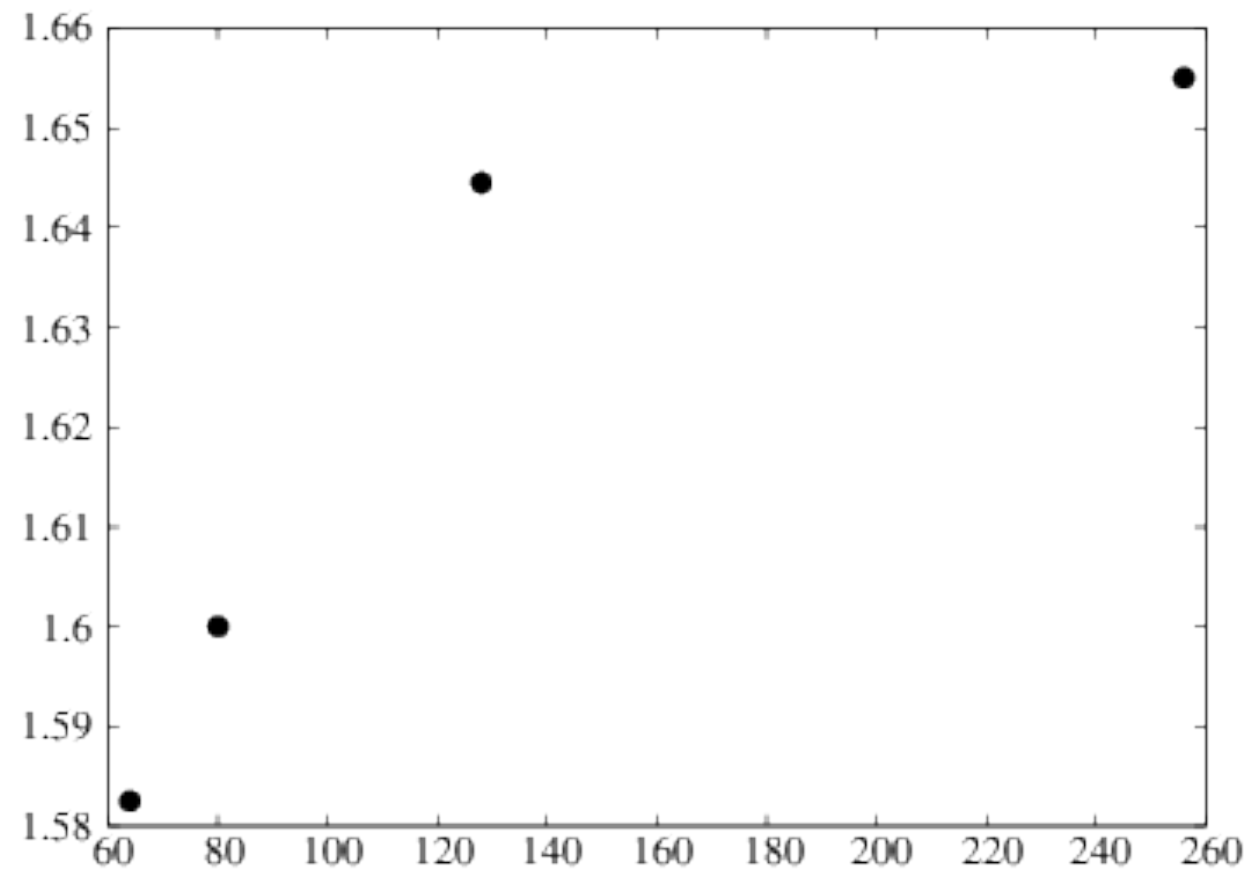}
\caption{The plot of current density (in vicinity of R as marked in panel b of Fig. \ref{initialrec}) 
against grid resolution. The abscissa is grid resolution along $z$ whereas the ordinate is current density.
The monotonous increase of current density with resolution asserts the 
increase of gradient in magnetic field. } \label{scaling}
\end{figure}

\clearpage

\begin{figure}
\centering
\includegraphics[angle=0,scale=.35]{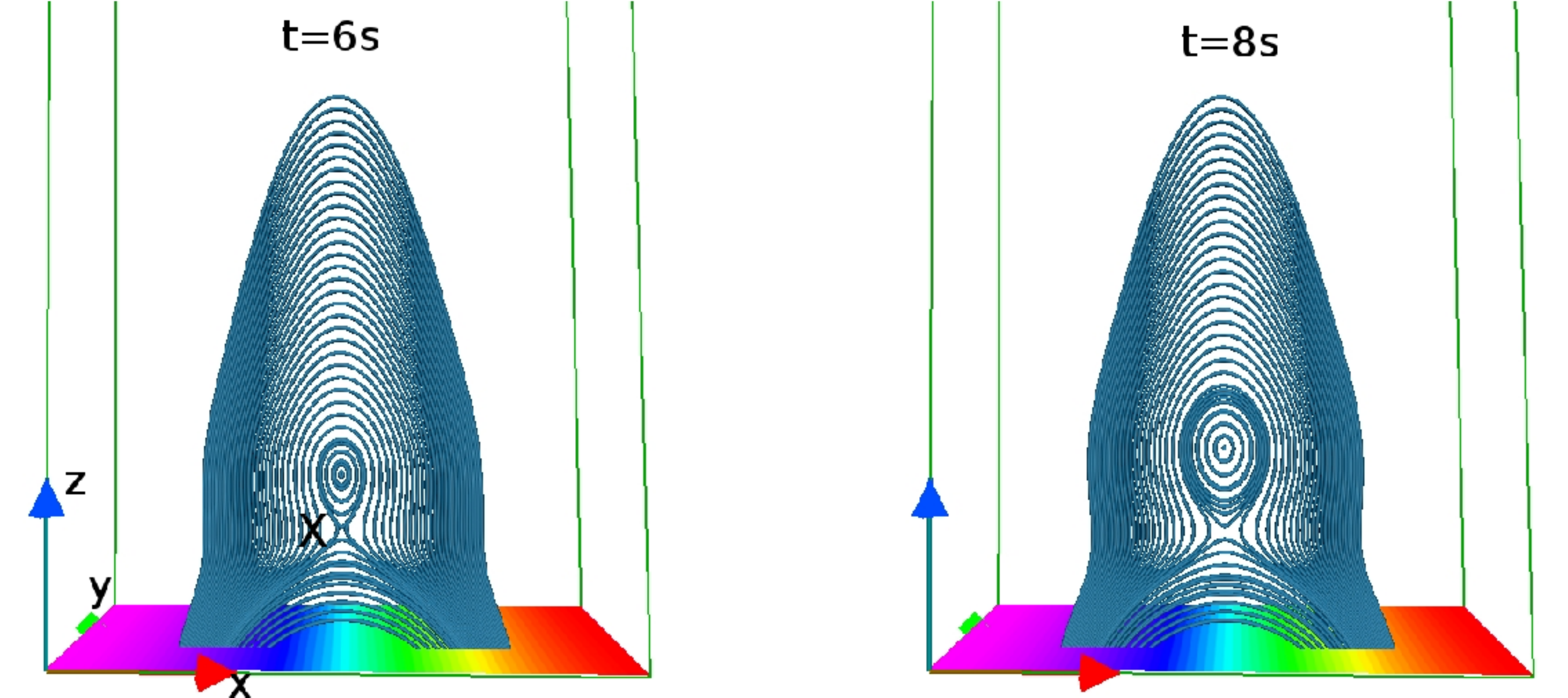}
\caption{Time sequence of field lines at instances $t=6s$ and $t=8s$, projected on a $y$-constant plane. 
Panel a documents the presence of a magnetic island, reminiscent of the rope,  
along with a $X$-type null (marked by the symbol X) below the island. 
The panel b shows the number of field lines constituting the island increases 
while the center rises along the vertical.} \label{xtype}
\end{figure}

\clearpage

\begin{figure}
\centering
\includegraphics[angle=0,scale=.35]{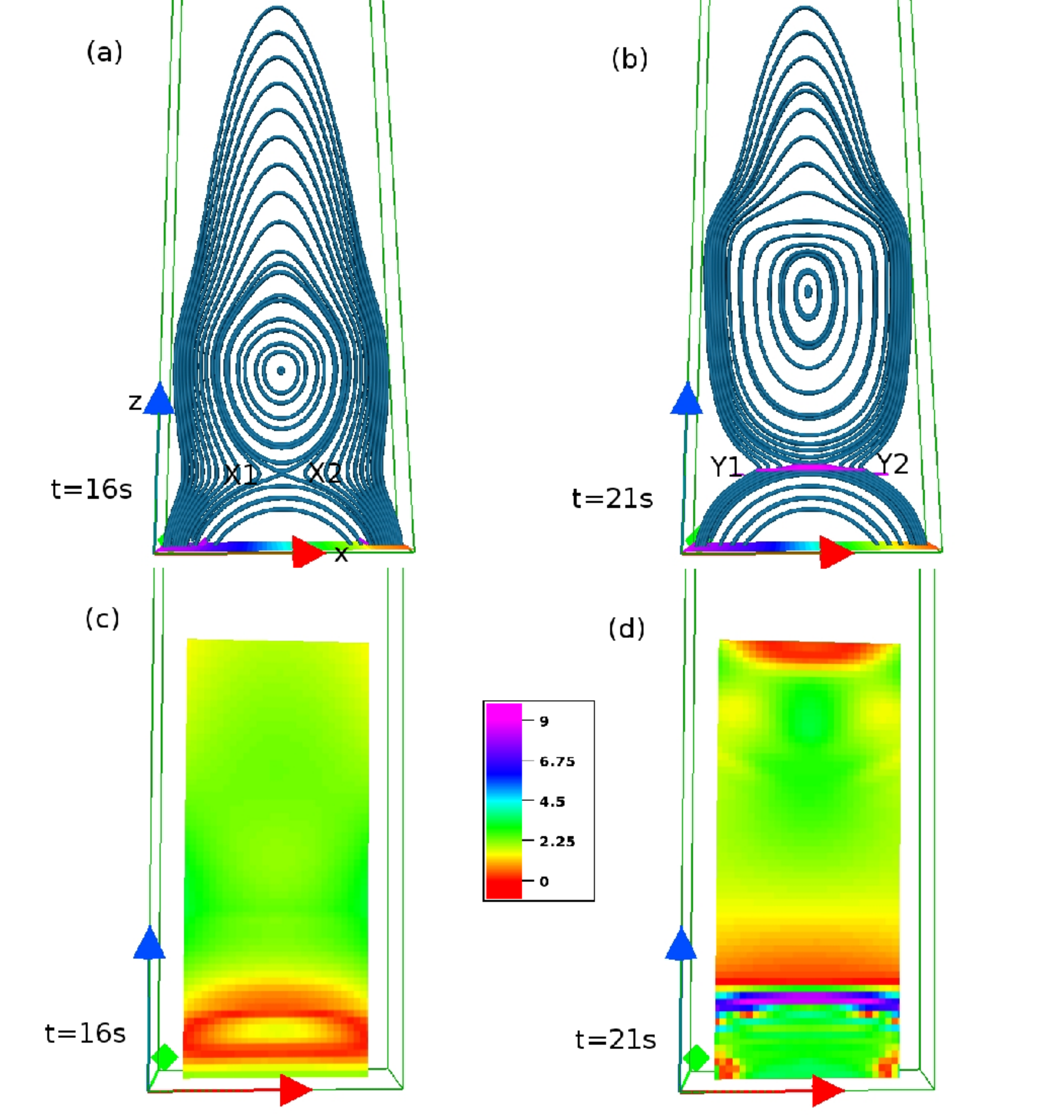}
\caption{Evolution of field lines (projected on a $y$-constant plane) coincides with quasi-steady state of the relaxation 
(panels a and b). In addition, the contours of magnitude of current density the $y$-constant plane 
are plotted in panels c and d. Noteworthy are the pressing of two quadrants 
(shown by symbols X1 and X2 in panel a) and generation of two Y-type nulls (denoted by symbols Y1 and Y2 in panel b) 
along with an extended CS (depicted in color pink in panel b). The development of 
extended CS is also attested by the current contours in panel d. } \label{extndcs}
\end{figure}

\clearpage

\begin{figure}
\centering
\includegraphics[angle=0,scale=.35]{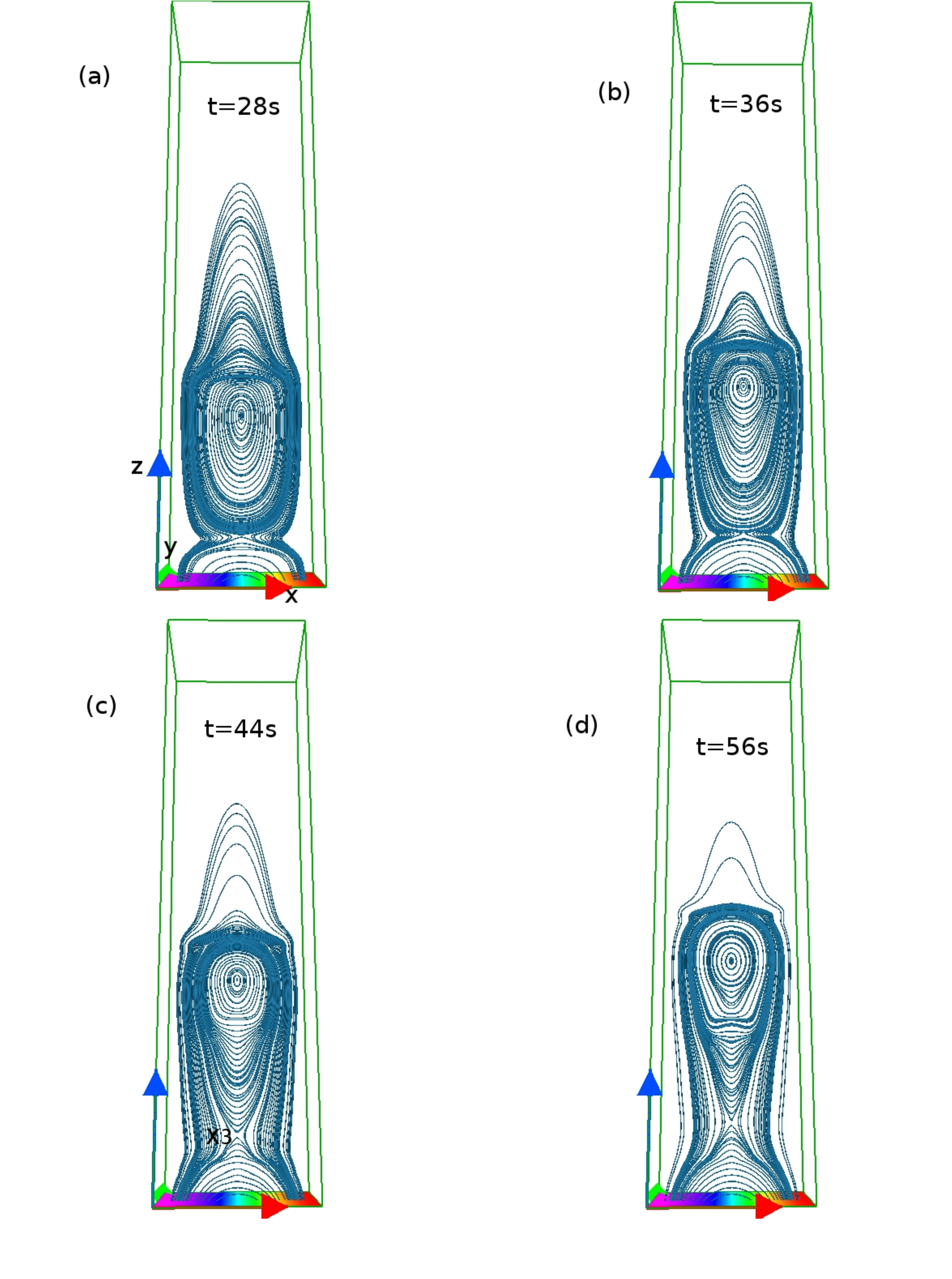}
\caption{The figure plots field lines during the third phase of evolution. The field lines are projected on 
a $y$-constant plane. Important is the onset of a new $X$-type null (illustrated by $X3$) while 
the bottom portion of the rope develops a dip (panel c).
Also, the newly formed $X$-point moves upward along with the rope (panel d).  } \label{ascend}
\end{figure}

\clearpage
\begin{figure}
\centering
\includegraphics[angle=0,scale=.40]{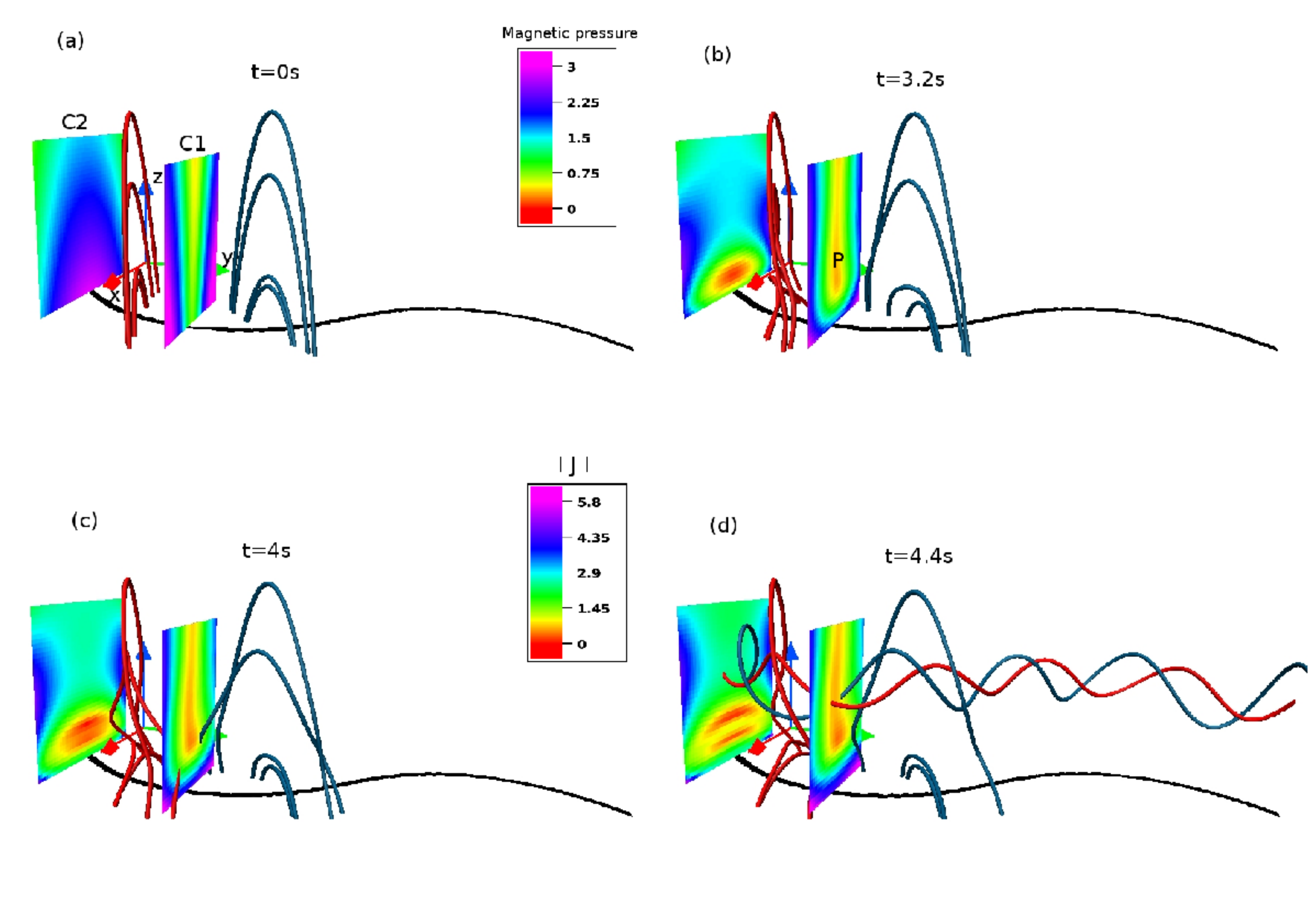}
\caption{Time sequences of two sets of field lines for the three dimensional simulation with 
initial field ${\bf{B}}^{\star}$.  
The panel a shows the field lines of the initial field.  
Important to note is the curved PIL (the solid black line).
The figure is further overlaid with contours of magnetic pressure (denoted by C1 in panel a) 
and current density (denoted by C2 in panel a) plotted
on different $y$-constant planes.  Panels b, c and d document the development of magnetic pressure depleted region 
(symbolized by P) in the $y$-constant plane and subsequent reconnection which leads to 
the generation of a flux-rope. Also, the flux-rope is situated above the PIL. Moreover, the 
current contours depict that no extended CS originates during this phase of the evolution.} \label{3dinitialrec}
\end{figure}

\clearpage
\begin{figure}
\centering
\includegraphics[angle=0,scale=.35]{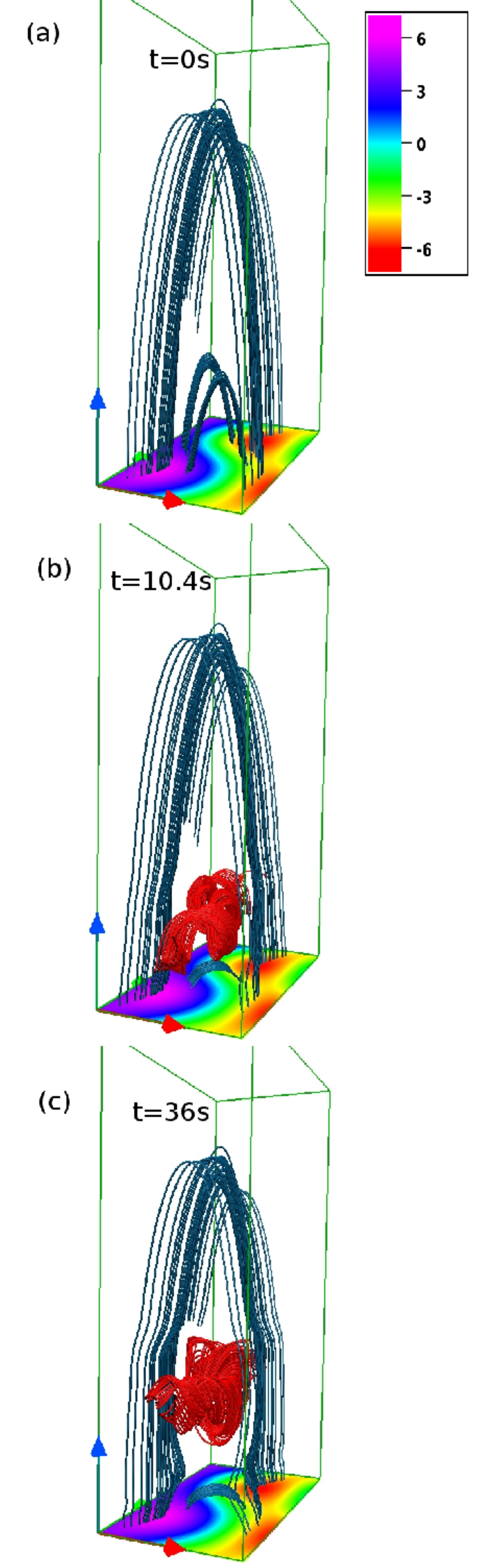}
\caption{Time sequences of evolution with more densely plotted field lines of the ${\bf{B}}^{\star}$. 
The lines in red marks the flux-rope. 
The overall evolution is similar to the 2.5D case. Noticeable is the ascend of the rope. 
The animated evolution is presented as supplementary material.} \label{3dmov}
\end{figure}

\clearpage
\begin{figure}
\centering
\includegraphics[angle=0,scale=.35]{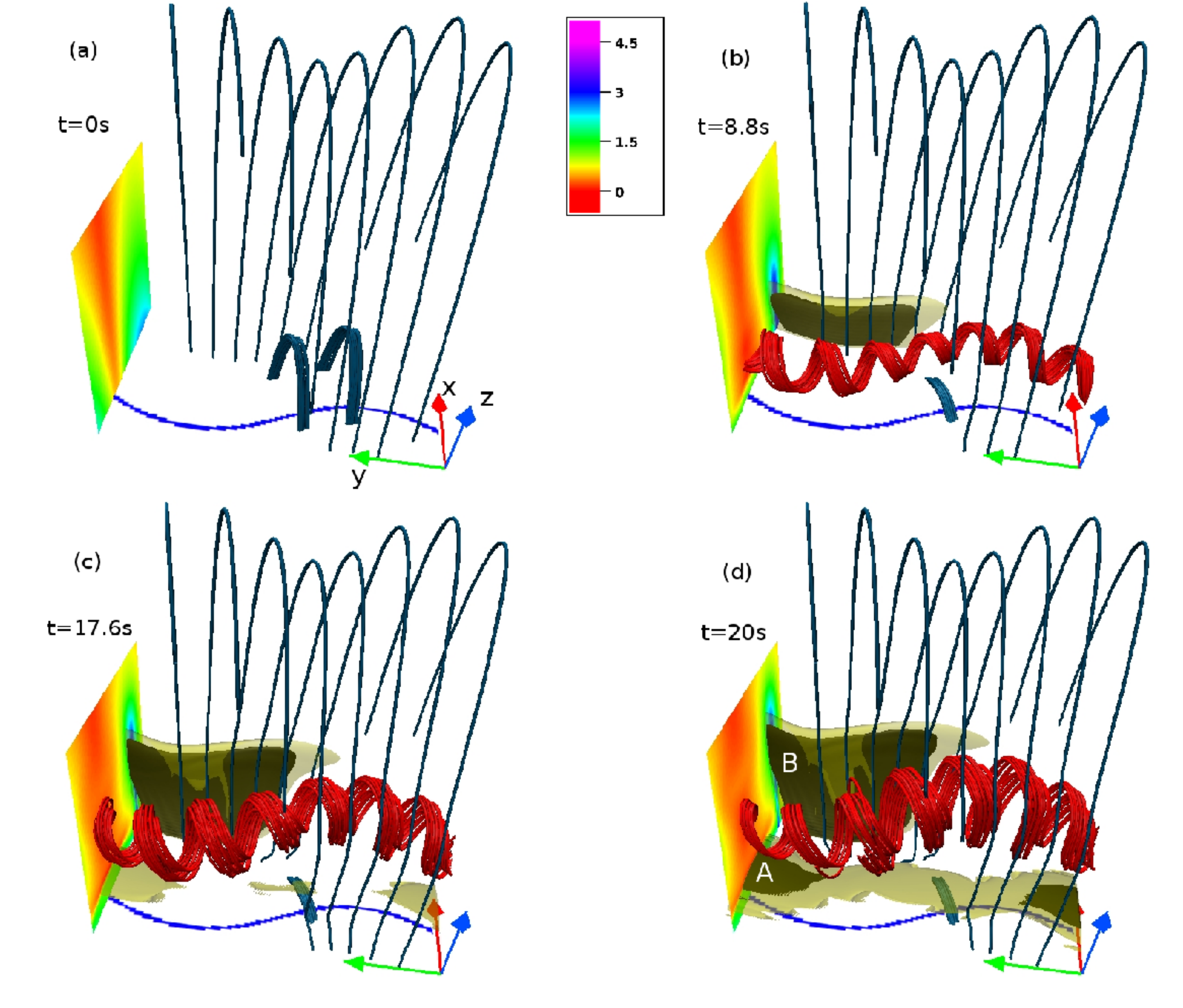}
\caption{Evolution of field lines for the three dimensional simulation with initial 
field ${\bf{B}}^{\star\star}$, illustrated in panel a (the curved PIL is shown by the 
solid blue line). The figure is further overlaid with isosurfaces of current density 
with isovalues $15\%$ (in color yellow) and $20\%$ (in color black) of its maximum 
and contours of $\mid{\bf{B}}^{\star\star}\mid$ on a $y$-constant plane. The helical 
red field lines identify the flux-rope which rises in the vertical direction. Notable
is the development of current sheet below the rope (marked by A in panel d). The elongated 
surfaces (marked by B in panel d) being co-located with enhanced $\mid{\bf{B}}^{\star\star}\mid$
region in $y$-constant plane imply that their onset does not indicate CS formation. The 
animated evolution is presented as supplementary material.} \label{fluxrope3}
\end{figure}

\end{document}